\documentclass[%
 aip,
 amsmath,amssymb,
preprint,%
]{revtex4-1}

\usepackage{graphicx}
\usepackage{dcolumn}
\usepackage{bm}

\usepackage{color}

\usepackage[utf8]{inputenc}
\usepackage[T1]{fontenc}
\usepackage{mathptmx}

\usepackage{color}
\usepackage{tikz}
\usepackage{lipsum}
\usepackage{soul}
\usepackage{cancel}
\usepackage{multirow}

\begin{document}

\def\beq{\begin{eqnarray}}
\def\eeq{\end{eqnarray}}
\newcommand{\stc}[1]{\textcolor{magenta}{\st{#1}}}
\newcommand{\tc}[1]{\textcolor{magenta}{#1}}

\def\O{{\cal O}}
\def\scal{\mathfrak{s}}
\def\N{{\bf N}}
\def\enu{\epsilon_\nu}
\def\hx_{\hath}
\def\hatx_{{\bf \hat x}}
\def\d{{\rm d}}
\def\im{{\rm i}}
\def\e{{\bf e}}
\def\ex{{\rm e}}
\def\x{{\bf x}}
\def\v{{\bf v}}
\def\X{{\bf X}}
\def\L{{\rm L}}
\def\k{{\bf k}}
\def\S{{\bf S}}
\def\J{{\bf J}}
\def\y{{\bf y}}
\def\t{{\bf t}}
\def\btau{{\boldsymbol{\tau}}}
\def\z{{\bf z}}
\def\R{{\bf R}}
\def\A{{\bf A}}
\def\r{{\bf r}}
\def\p{{\bf p}}
\def\u{{\bf u}}
\def\n{{\bf n}}
\def\U{{\bf U}}
\def\V{{\bf V}}
\def\W{{\bf W}}
\def\F{{\bf F}}
\def\T{{\bf T}}
\def\cm{{\rm cm}}
\def\q{{\bf q}}
\def\sec{{\rm sec}}
\def\Ckol{C_{Kol}}
\def\flux_{\bar\epsilon}
\def\b{b_{kpq}}
\def\bchi{{\boldsymbol{\chi}}}
\def\bOmega{{\bf\Omega}}
\def\balpha{{\boldsymbol{\alpha}}}
\def\nf{{\scriptscriptstyle nf}}
\def\smalfi{{\scriptscriptstyle \frac{5}{3} }}
\def\smalV{{\scriptscriptstyle {V}}}
\def\ice{{\scriptscriptstyle {ice}}}
\def\smalWt{{\scriptscriptstyle \tilde{\rm W}}}
\def\smalWWt{{\scriptscriptstyle {\rm W}[\tilde{\rm W}]}}
\def\smalT{{\scriptscriptstyle {\rm T}}}
\def\smalE{{\scriptscriptstyle{\rm E}}}
\def\smaln{{\scriptscriptstyle (n)}}
\def\smalnm1{{\scriptscriptstyle (n-1)}}
\def\smalnp1{{\scriptscriptstyle (n+1)}}
\def\smalnm2{{\scriptscriptstyle (n-2)}}
\def\smalze{{\scriptscriptstyle (0)}}
\def\smalun{{\scriptscriptstyle (1)}}
\def\smaldu{{\scriptscriptstyle (2)}}
\def\smalk{{\scriptscriptstyle k}}
\def\smalel{{\scriptscriptstyle l}}
\def\gammaP{\gamma^\smalP}
\def\shell{{\tt S}}
\def\ball{{\tt B}}
\def\nav{\bar N}
\def\micron{\mu{\rm m}}
\font\brm=cmr10 at 24truept
\font\bfm=cmbx10 at 15truept

\preprint{AIP/123-QED}

\title[Diffusion of gravity waves by random space inhomogeneities
in pancake-ice fields.]{Diffusion of gravity waves by random space inhomogeneities
in pancake-ice fields. \\
Theory and validation with wave buoys and SAR}

\author{P. Olla}
 \altaffiliation[Also at ]{INFN, Sez. Cagliari, I--09042 Monserrato, Italy}
 \email{olla@dsf.unica.it}
\affiliation{ 
ISAC-CNR Sez. Cagliari, I--09042 Monserrato, Italy
}

\author{G. De Carolis}
\author{F. De Santi}
\affiliation{%
IREA-CNR, Sez. Milano,  I--20133 Milano, Italy}%

\date{\today}

\begin{abstract}
We study the diffusion of ocean waves by ice bodies much smaller than a wavelength, such as pancakes and small ice floes. We argue that
inhomogeneities in the ice cover at scales comparable to that of the wavelength significantly increase diffusion, producing a contribution to wave attenuation
comparable to what it is observed in the field and usually explained by viscous effects.
The resulting attenuation spectrum is characterized by a peak at the scale of the inhomogeneities in the ice cover, which  could explain the  rollover of the attenuation profile at small wavelengths observed in field experiments. 
The proposed attenuation mechanism leads to the same behaviors that would be produced by a viscous wave model with effective viscosity linearly dependent on the ice thickness. This may explain recent findings that viscous wave models require a thickness-dependent viscosity to fit experimental attenuation data.
Experimental validation is carried out using wave buoy attenuation data and SAR image analysis.

\end{abstract}

\maketitle



\section{Introduction}
Ocean surface waves play an important role in the dynamics of sea ice in polar regions. The effect is particularly visible in the Marginal Ice Zone (MIZ);, which is the highly dynamical transition region separating the ice pack from the open ocean\cite{squire95}. In the summer season, erosion of the pack and breakup of large flows into smaller ice bodies, are the dominant processes in the region. 
During winter, waves infiltrating the MIZ contribute to ice formation through the coalescence of ice bodies into progressively larger objects \cite{thomson2014swell}.
Global warming has led to an increase in both the activity and the extension of the MIZ.
The Arctic in particular has seen a dramatic summertime increase of the MIZ extension and a
concomitant reduction of the ice 
surface \cite{parkinson13,thomson18}.
To determine how deep ocean waves can propagate into the MIZ, one must know the sea ice contribution to wave damping. Over the years, a great effort has been put into
studyng the modifications in surface waves propagation induced by sea ice (see Squire \cite{squire95,squire07,squire20} for an extended review).
In particular, remote sensing analysis of the modifications to wave propagation by sea ice has been used as a tool to infer the ice thickness without having to resort to in situ measurements \cite{wadhams1999mapping, de2001retrieval,  wadhams02, decarolis2003sar, wadhams2004sar, sutherland2016airborne, wadhams2018pancake, decarolis21}.

Wave attenuation depends on the characteristic
size of the ice bodies involved. In the case
of larger floes on the scale of tens to hundreds of meters, scattering and diffraction \cite{masson89,bennetts12}, as well as flexural deformations of the floes \cite{meylan96,meylan2015experimental}, are expected to dominate (see \citet{li2021interaction} for a
recent application of thin plate theory to wave propagation in water covered by a continuous ice sheet). On the other hand, in the case of smaller ice bodies, viscous effects 
and collisions are expected to be dominant.
 \cite{kohout08,squire16}.
The picture is complicated by the strong inhomogeneity of the MIZ; 
although small ice is prevalent at the fringes of the MIZ, ice bodies
of widely different sizes and typology can be found anywhere in the region \cite{toyota11}. 

The focus of the present paper is the modification of ocean wave propagation produced
by ice bodies much smaller than a wavelength, in particular pancake-ice.
Pancake-ice is the assembly of pancake-shaped ice bodies of 
diameters ranging from 30 centimeters to 3 meters, which
populate, during ice formation,
the outer fringes of the MIZ \cite{doble03,doble15}.
It usually comes as a component of so-called grease-pancake ice (GPI),
i.e. pancake-ice embedded in a dense slurry of ice crystals, called grease ice. 

Grease ice is a strongly viscous medium, which 
is usually treated as a viscous continuum \cite{keller98}. Over the years various extensions
of the viscous model have been proposed, such as the inclusion of 
viscoelasticity in the ice response
\cite{wang10,chen19} and of turbulent effects 
in the water under the ice \cite{decarolis02}, and the treatment of pancakes as an 
individual layer separate from grease ice \cite{desanti17}.

Providing a microscopic explanation 
for the effective viscosity of GPI, however, has proven
problematic. This is reflected in the wide variation of the values of the parameter 
required to fit different experimental data sets \cite{cheng2017calibrating,de2018ocean}.
The difficulty is compounded by recent observations \cite{decarolis21,sutherland19,doble15,li2017rollover} that most viscous layer models require an effective viscosity dependent on the ice thickness to fit available experimental data. 
The possibility that an intensive quantity such as the ice viscosity depends
on a macroscopic quantity such as the ice layer thickness casts doubts on whether 
 GPI could be treated as a simple viscous medium.
 Besides, there are situations in which pancakes (or other forms of multi-year small scale or brash ice) float in low-viscosity grease-ice-free water \cite{thomson18}.

Another difficulty is that current viscous models are unable to explain the rollover effect, i.e. the decrease of wave attenuation at short wavelengths which is often observed in field measurements. Although this phenomenon has been discussed for over four decades, it still lacks of an exhaustive explanation.
The rollover effect has been so far
ascribed to instrumental noise \cite{thomson2020spurious}, to exogenous mechanisms not properly taken into account, such as the energy input by the wind \cite{li2017rollover}, and to the effect of nonlinearities \cite{polnikov2007calculation}. 
A possible explanation of the phenomenon was proposed by \citet{liu1988wave} as the result of the development of a
viscous boundary layer under a continuous ice sheet treated as a thin elastic plate.
Another explanation of the phenomenon was proposed by
\citet{perrie1996air},
who performed numerical simulations with constant wind and regular arrays of medium size circular ice floes (diameter $20\,{\rm m}$ ), suggesting that rollover is indeed the result of wave scattering.
To our knowledge, however, scattering theory has never been applied at pancake scales yet.

All these remarks call for a reappraisal of alternative mechanisms of wave attenuation, such as, e.g., collisions between
ice bodies \cite{shen98}, nonlinear forces and turbulence induced in the surrounding water by the ice motion \cite{skene15}.

Purpose of the present study is to evaluate the  scattering contribution to wave attenuation. 
We show that
random inhomogeneities in the ice cover generate a coherent scattering component that can contribute significantly to wave damping. The phenomenon is akin to the enhancement of light scattering in a turbid medium. 
The approach we are going to follow is common to that for larger floes, 
and is based on partial wave expansion of the flow disturbances generated by the inhomogeneities of the ice cover \cite{kagemoto86}. The nature of the scattering process in the
case of floating bodies much smaller than a wavelength, however, is fundamentally different,
as bodies within a wavelength respond coherently to the wave, and subtle
near-field effects simplify the treatment of the scattering dynamics. 

We assume potential flow for both the incident wave and the disturbances,
thus disregarding viscous effects from the possible presence of grease ice at the water surface.
Small amplitude waves and linear hydrodynamics
are assumed throughout the analysis; 
for simplicity, only the case of deep water waves is considered.
We treat the ice layer as an inextensible continuum with vanishing resistance to bending
and with a rough bottom. In the case of a spatially uniform layer and no roughness, the dynamics reduce to that of the mass-loading model \cite{peters50,keller53}.

We compare our theoretical results with experimental data from wave buoys collected during the Sikuliaq campaign in Autumn 2015 \cite{thomson18}. By exploiting the availability of attendant satellite SAR images available, it is possible to estimate the scale of the ice cover spatial inhomogeneities, which appears to coincide with that of the rollover peak.

The paper is organized as follows. In Sec. II, the general theory of scattering of gravity waves by floating bodies is reviewed. In Secs. III and IV, we apply the theory to determine the modification
to the wavefield generated by random spatial variations in a continuous cover. In Sec. V, the consequences of the results on wave attenuation are discussed. In Sec. VI, the specific example of a spatially inhomogeneous mass loading with a rough ice-water interface, is considered. In Section VII, current results are contextualized to the others wave propagation models in the literature.  Section VIII contains a comparison with experimental data.
Concluding remarks are reported in Section IX.

\section{Partial wave expansion}
\label{Partial wave}
The study of the scattering of ocean waves by solid obstacles has a long history, which makes
extensive use of techniques drawn from the study of similar problems in electromagnetism,
acoustics and quantum mechanics. 
The approach has been detailed elsewhere (see e.g. \citet{peter04a}). We
outline here the essentials.

The wave dynamics obeys the linearized Euler equation 
\beq
\rho\partial_t\U+\nabla P=0,
\quad
\nabla\cdot\U=0,
\label{Euler}
\eeq
where $\U(\x,t)$ is the fluid velocity, 
$\rho\simeq 1025\ {\rm kg\,m^{-3}}$ is the 
salt water density, $P(\x,t)$ is the deviation of the pressure in the water column
from its value at hydrostatic equilibrium,
and we have put the unperturbed water surface at $x_3=0$ 
with $\e_3$ pointing upwards.
Equation (\ref{Euler}) implies potential flow
\beq
\U=-\nabla\Phi,
\qquad
\nabla^2\Phi=0.
\label{potential}
\eeq
Continuity of the pressure at the water surface, $P(\x(t),t)|_{x_3=0}=0$, where
$\x(t)$ is the instantaneous position of the point at
the water-atmosphere interface, 
leads to the boundary condition on Eq.
(\ref{potential})\cite{landauFM}
\beq
\Big(\frac{\partial^2\Phi}{\partial t^2}+g\frac{\partial\Phi}{\partial x_3}\Big)_{x_3=0}=0.
\label{water surface}
\eeq

The wave energy contains both potential and kinetic energy contributions.
We can eliminate the potential energy component by averaging over a suitable time window
$w$, and thus get the expression for the time averaged wave energy density
\beq
\mathcal{E}_\Phi(\x_\perp,t)&=&
\frac{\rho}{2}\int_{-\infty}^0\d x_3\ \langle|\U(\x,t)|^2\rangle_w,
\label{E_Phi}
\eeq
where $\x=(\x_\perp,x_3)$ and $\langle.\rangle_w$ indicates time average in $w$.
By comparing Eq. (\ref{E_Phi}) with
Eqs. (\ref{potential}) and (\ref{water surface}) we find 
the conservation law
\beq
\partial_t\mathcal{E}_\Phi+\nabla_\perp\cdot\J_\Phi=0,
\qquad
\J_\Phi(\x_\perp,t)=\int_{-\infty}^0\d x_3\ \langle P(\x,t)\U_\perp(\x,t)\rangle_w.
\eeq
We assume elastic scattering, which implies linearity of possible mechanical interactions
of the floating bodies;
linear hydrodynamics then implies that the scattered
waves oscillate with the same frequency of the incident wave. In the case
of a monocromatic incident wave of frequency $\omega$, the energy current
$\J_\Phi$ will satisfy
\beq
\J_\Phi=
\im\rho\omega\int_{-\infty}^0\d x_3\ [\Phi\nabla_\perp\Phi^*-\Phi^*\nabla_\perp\Phi].
\label{J}
\eeq

We are interested in the scattering by ice bodies of characteristic size $R$ much less
than a wavelength.
Taking inspiration from the treatment of similar problems in the case of electromagnetic 
radiation \cite{jackson}, we separate in the potential a near-field component $\phi$
from disturbances from bodies at distance $\ll k^{-1}$ from $\x$, and a remnant $\Psi$ containing the incident and the diffused waves:
\beq
\Phi(\x,t)=\Psi(\x,t)+\phi(\x,t).
\label{decompose 1}
\eeq
A summary of the relevant scales in the problem is provided in Table \ref{table1}.
\begin{table}[h]
\begin{center}
\caption{Relevant scales of the problem}
\label{table1}
\begin{tabular}{c|lc}
\toprule
\multirow{2}{*}{Near field $kL\ll 1$} & Pancake scale $L\sim R$. & Pancake geometry important.
\\
&Intermediate scale $L\gg R$. & Multipole expansion 
approach.
\\
\hline
Wave region $kL\sim 1$ & Perturbations acquire a wave character ---&assumed scale of
ice inhomogeneities ($\lambda$).
\\
\hline
Far field $kL\gg 1$ & Diffused waves. &
\\
\hline\hline
\end{tabular}
\end{center}
\end{table}

We are interested in the scattering by random space inhomogeneity of the ice cover.
Let us then indicate with overbar 
the operation of ensemble average with respect to the homogeneities of the ice cover
and with tilde fluctuation, and decompose the generic physical quantity $f$ accordingly:
\beq
f=\bar f+\tilde f.
\eeq
Hence $\bar\Psi$ contains the incident wave, $\tilde\Psi$ contains the
diffused waves; ensemble averaging and coarse-graining have the same filtering
effect on the small scales, therefore $\bar\Phi=\bar\Psi$ and $\phi=\tilde\phi$.
\begin{figure}[ht]
    \centering
\includegraphics[width=8cm]{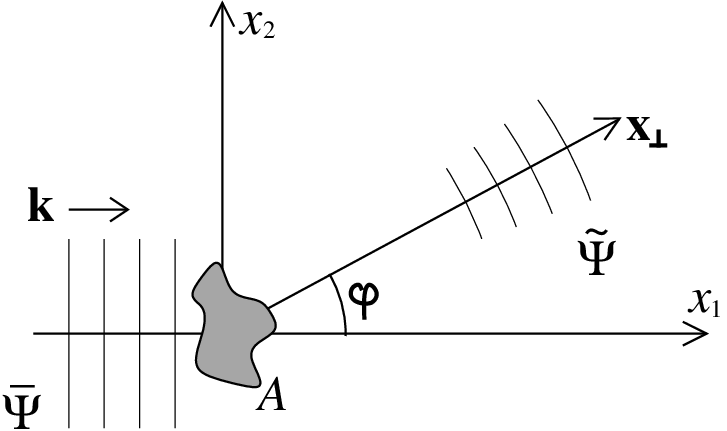}
\caption{Geometry of the problem}
    \label{radfig0}
\end{figure}
Consider a portion $A$ of unperturbed water surface at $\x_\perp=0$
and indicate with $\tilde\Psi_A$ the contribution 
to wave diffusion from the bodies in that region.The geometry of the problem is sketched in Fig. \ref{radfig0}.
Let us assume a monocromatic incident wave propagating to positive $x_1$: $\k=k\e_1$;
from Eqs. (\ref{potential}) and (\ref{water surface}) we then get the classical deep-water surface gravity wave solution in ice-free water
\beq
\bar\Psi(\x,t)=\bar\Psi_0(t)\exp(kx_3+\im kx_1)=\bar\Psi_0\exp[kx_3+\im(kx_1-\omega t)],
\quad\omega^2=kg.
\eeq
We can expand the incident wave in partial waves
\beq
\bar\Psi(\x,t)=
\bar\Psi(t)\Big[J_0(kx_\perp)
+2\sum_{n=1}^\infty\im^nJ_n(kx_\perp)\cos(n\varphi)\Big]\ex^{kx_3},
\label{Phi_inc}
\eeq 
and similarly for $\Psi_A=\bar\Psi+\tilde\Psi_A$,
\beq
\Psi_A(\x,t)&=&\tilde\Psi_{A,eva}(\x,t)+\bar\Psi_0(t)\Big[B_0J_0(kx_\perp)+C_0N_0(kx_\perp)
\nonumber
\\
&+&2\sum_{n=1}^\infty\im^n\Big(B_nJ_n(kx_\perp)+\im C_n
N_n(kx_\perp)\Big)\cos(n\varphi)\Big]\ex^{kx_3},
\label{Phi_eva}
\eeq
where $\cos\varphi=x_1/x_\perp$, $J_n$ and $N_n$ are Bessel functions of the first and second
kind, respectively,
and $\tilde\Psi_{A,eva}$ contains evanescent modes, which 
can be expressed in terms of modified Bessel functions $K_n$,
exponentially decaying at large $kx_\perp$ \cite{abramowitz}. 
The evanescent modes can 
be disregarded in the far-field region, 
and the remaining part of $\Psi_A$ is a sum of outward and inward 
propagating modes $\propto\ex^{\pm\im kx_\perp}$. Requiring $\tilde\Psi_A$
not to contain inward propagating modes (Sommerfeld condition), together
with stationarity, $\J_{\bar\Psi}=0$, 
allows us to fix the coefficients in Eq. (\ref{Phi_eva}) \cite{landauQM}. 
This gives us the expression
for the perturbed field in the far-field region $kx_\perp\gg 1$
\beq
\tilde\Psi_A(\x,t)&=&\bar\Psi_0(t)\Big[\frac{\ex^{2\im \delta_{A,0}-1}}{2}H_0^\smalun(kx_\perp)
+\sum_{n=1}^\infty\im^n
(\ex^{2\im\delta_{A,n}}-1)
\nonumber
\\
&\times&
H_n^\smalun(kx_\perp)
\cos(n\varphi)\Big]\ex^{kx_3},
\qquad\quad kx_\perp\gg 1,
\label{phi_out 0}
\eeq
where $H^\smalun_n=J_n+\im N_n$ are Hankel function of the first kind \cite{abramowitz},
and the coefficients $\delta_{A,n}$ are real constants called the scattering phases.

We confine our attention to situations in which scattering  can be considered
a perturbation; the scattering phases therefore are small and we can
linearize Eq. (\ref{phi_out 0}):
\beq
\tilde\Psi_A(\x,t)&\equiv&\sum_n\tilde\Psi_{A,n}(\x,t)
\simeq\im\bar\Psi_0(t)\Big[\delta_{A,0}H_0^\smalun(kx_\perp)
\nonumber
\\
&+&2\sum_{n=1}^\infty\im^n
\delta_{A,n}H_n^\smalun(kx_\perp)
\cos(n\varphi)\Big]\ex^{kx_3},
\qquad kx_\perp\gg 1.
\label{phi_out}
\eeq
Let us suppose that $A$ contains all the bodies responsible for scattering,  
or equivalently that the interference with waves scattered by bodies
outside $A$ is negligible. 
In this case, the radiated energy is 
\beq
I_{\tilde\Psi_A}=\int_0^{2\pi}\d\varphi\ \x_\perp\cdot\J_{\tilde\Psi_A}(\x),
\label{I_phi}
\eeq
where $J_{\tilde\Psi_A}$ is obtained from Eq. (\ref{J}) 
by replacing $\Phi$ with $\tilde\Psi_A$.
Substituting Eq. (\ref{phi_out}) into Eq. (\ref{I_phi})
and exploiting the relation $N'_n(z)J_n(z)-N_n(z)J'_n(z)=2/(\pi z)$ \cite{abramowitz}
yields
\beq
I_{\tilde\Psi_A}\equiv\sum_nI_{\tilde\Psi_{A,n}}=
\int_0^{2\pi}\d\varphi\ \x_\perp\cdot\J_{\tilde\Psi_A}(\x)
\simeq\frac{4\rho\omega|\bar\Psi_0|^2}{k}\sum_{n=0}^\infty\delta^2_{A,n}.
\label{I}
\eeq

By combining Eq. (\ref{I}) with the energy flux density of the incident wave
\beq
\J_{\bar\Psi}=\omega\rho|\bar\Psi_0|^2\e_1,
\eeq
we obtain the expression for the scattering cross-section
\beq
\sigma_A=\frac{I_{\tilde\Psi_A}}{J_{\bar\Psi}}\simeq 4k^{-1}\sum_{n=0}^\infty\delta_{A,n}^2.
\label{sigma}
\eeq

\section{The near field}
\label{Near field}
Let us consider a region $A$ at the water surface around $x_\perp=0$, of size $L_A$
sufficiently large to carry out a continuous limit,
but small compared to the wavelength: $R\ll L_A\ll k^{-1}$.
The near-field disturbance by the bodies in $A$
is a complicated superposition of propagating and evanescent
modes \cite{kagemoto86,peter04}, which is more aptly analyzed by making the decomposition
$\phi_A=\phi_{A,reg}+\phi_{A,sing}$, where 
$\phi_{A,reg}$ and $\phi_{A,sing}$, are the regular (singular) components of $\phi$ at
$\x_\perp=0$, $R,L\to 0$, and both components   
are required to decay for $x_\perp\to\infty$.

The regular component can only contain
Bessel functions $J_n$. We therefore obtain from Eq. (\ref{phi_out}) the expression,
valid for all $kx_\perp$,
\beq
\phi_{A,reg}\equiv\sum_n\phi_{A,reg,n}\simeq\im\bar\Psi_0\Big[\delta_0J_0(kx_\perp)
+2\sum_{n=1}^\infty\im^n
\delta_nJ_n(kx_\perp)
\cos(n\varphi)\Big]\ex^{kx_3}.
\label{phi_slow}
\eeq

The singular component $\phi_{A,sing}$ is a superposition of Bessel functions
$N_n$ and $K_n$, out of which, only the $N_n$ component survives in the far-field region $kx_\perp\gg 1$, 
where it forms the imaginary part of the Hankel functions in Eq. (\ref{phi_out}).
Let us focus on an intermediate region $L_A\ll x_\perp\ll k^{-1}$ in the near field. We assume that the hydrodynamics interaction of the bodies decays sufficiently fast with their separation. In a first approximation, the perturbed potential 
$\phi_{A,sing}$ can therefore be evaluated at $x_\perp\gg L_A$ using the ice-free boundary condition in Eq. (\ref{water surface}).
In that region, $\phi_{A,sing}$ sees $A$ as a point and is expected 
to vary at scale $x$; since the time scale of wave
disturbances at scales $x_\perp\ll k^{-1}$ is much shorter than $\omega^{-1}$, we can then
replace the Robin boundary condition in Eq. (\ref{water surface}) with 
the Neumann boundary condition 
\beq
\partial_{x_3}\phi_{A,sing}(\x,t)|_{x_3}=0,
\qquad
x_\perp\ne 0.
\eeq
The perturbed potential
$\phi_{A,sing}$ will thus behave in the intermediate region like 
the electrostatic potential of a superposition of multipoles at $\x=0$.
This allows us to do away with the evaluation of
evanescent and propagating modes in terms of Bessel functions.
Linearity of the dynamics allows us to write
the perturbed potential in the form
\beq
\phi_{A,sing}(\x,t)\equiv\sum_n\phi_{A,sing,n}(\x,t)\simeq
\frac{1}{2\pi}\Big[\frac{\alpha_{A,0}}{x}
+\frac{\balpha_{A,1}\cdot\x_\perp}{x^3}+
\frac{\balpha_{A,2}:\x_\perp\x_\perp}{x^5}
+\ldots\Big]
\bar\Psi_0(\x,t),
\label{phi_in,f}
\eeq
where the $\balpha_{A,n}$'s are $n$-index symmetric zero-trace tensors 
playing a role analogous to the polarizability of a small body in the field of an electromagnetic 
wave, which enter Eq. (\ref{phi_in,f})  
 contracted with $n$ terms $\x_\perp'$ and  with a weigh factor $x^{-1-2n}$.

We can write Eq. (\ref{phi_in,f}) as a surface integral at $x'_3=0$:
\beq
\phi_{A,sing}(\x,t)&=&\frac{1}{2\pi}\int_A\d^2x_\perp'\,\bar\Psi(\x',t)\delta(\x'_\perp)[\alpha_{A,0}(\x'_\perp)
+\balpha_{A,1}(\x'_\perp)\cdot\nabla_\perp'+\ldots]|\x-\x'|^{-1}
\nonumber
\\
&=&\frac{1}{2\pi}\int_A\d^2x_\perp'\, |\x-\x'|^{-1}
[\alpha_{A,0}(\x'_\perp)-\nabla'_\perp\cdot\balpha_{A,1}(\x'_\perp)+\ldots]
\bar\Psi(\x',t)\delta(\x'_\perp),
\nonumber
\eeq
where in the second line of the equation the horizontal gradient $\nabla'_\perp$ acts both on $\balpha_{A,1}(\x')$ and on the factors $\bar\Psi(\x',t)\delta(\x_\perp')$ to the right of the square brackets.
The surface charge distribution 
$[\alpha_{A,0}(\x_\perp)-\nabla_\perp\cdot\balpha_{A,1}(\x_\perp)+\ldots]\bar\Psi(\x,t)\delta(\x_\perp)$
induces the boundary condition on the Laplace equation 
$\nabla^2\phi_{A,sing}=0$ \cite{jackson}:
\beq
\partial_{x_3}\phi_{A,sing}(\x,t)|_{x_3=0}=\Big[\alpha_{A,0}(\x_\perp)-
\nabla_\perp\cdot\balpha_{A,1}(\x_\perp)
+\ldots\Big]\bar\Psi(\x,t)\delta(\x_\perp).
\label{monopole}
\eeq

Equation (\ref{monopole}) tells us that the
effect of the bodies in $A$ is a localized perturbation of the vertical component of the 
fluid velocity at
the surface. The situation is particularly clear in the case of an isolated body.
If we take $A$ to coincide with the horizontal section of the body at $x_3=0$, 
$V_3=-\alpha_{A,0}\bar\Psi/A$ will be precisely
the vertical velocity of the body relative to the water surface.

The next contribution is a dipole, which accounts for the horizontal motion of the body
with respect to the wave. We have
\beq
\frac{\balpha_{A,1}\bar\Psi_0}{2\pi}=-\frac{1}{2\pi}\int_S\d S\,\x\partial_n\phi_{A,sing}(\x)=
\frac{1}{2\pi}\int_S\d S\,\x V_n(\x),
\label{dipole}
\eeq
where $\n(\x)$ is the normal at point $\x$ on the submerged part $S$ of the body surface
and $\V$ is the velocity of the body relative to the wave field. 
In the case of a fixed horizontal disk, such that
$\V=-\U$, we would have
\beq
\balpha_{A,1}\simeq \im\pi kR^2h\e_1,
\label{alpha_1}
\eeq
where $h$ and $R$ are the draft and the radius of the disk. 

Higher order multipoles take into account the effect of the non-uniformity of the 
velocity field of the wave at the scale of the body and involve additional factors
$kR$. For small $kR$ they can therefore be disregarded.

\section{The far field}
\label{Far field}
Let us now take for $A$ a region of size $L_A\gg\lambda$, where $\lambda$ is the 
correlation length of the fluctuations in the ice cover. 
Since we are assuming $\lambda$
comparable with the wavelength of the incident waves, the area element $A$ 
in Sec. \ref{Near field} must be considered infinitesimal compared to the one we are dealing with here. We introduce intensive quantities
\beq
\bchi_n=\frac{\d\balpha_n}{\d A}(\x_\perp)
\label{chi}
\eeq
giving the susceptibility of the medium.
We are assume local isotropy
of the ice cover, so that that $\bchi_1=\chi_1\e_1$, with similar expressions holding 
for higher-order multipoles.

At scales $x_\perp\gtrsim k^{-1}$, the time derivative
in the boundary condition Eq. (\ref{monopole}) must be restored:
\beq
(g\partial_{x_3}-\omega^2)\Psi|_{x_3=0}
\simeq g\chi\bar\Psi|_{x_3=0}\equiv[\chi_0-\nabla_\perp\cdot(\bchi_1\bar\Psi)
]_{x_3=0}.
\label{Robincoarse}
\eeq
Equations for the mean and the fluctuating components of $\Psi$ can be 
obtained from Eq. (\ref{Robincoarse}) by
following the strategy described in \citet{howe71}. We find
\beq
&&\nabla^2\bar\Psi=\nabla^2\tilde\Psi=0,
\\
&&\Big(g\partial_{x_3}-\omega^2\Big)\bar\Psi \Big|_{x_3=0}
=g\Big[\overline{\chi}\bar\Psi 
+\overline{\tilde\chi\tilde\Psi}\Big]_{x_3=0},
\label{mean0}
\\
&&\Big(g\partial_{x_3}-\omega^2\Big)\tilde\Psi\Big|_{x_3=0}
=g\Big[\tilde\chi\bar\Psi +\overline{\chi}\tilde\Psi
+\tilde\chi\tilde\Psi-\overline{\tilde\chi\tilde\Psi}\Big]_{x_3=0}.
\label{fluct0}
\eeq
For $\chi\ll 1$, Eqs. (\ref{mean0}) and (\ref{fluct0})  approximate to
\beq
&&\Big(g\partial_{x_3}-\omega^2\Big)\bar\Psi (\x,t)\Big|_{x_3=0}
\simeq g\overline{\chi}\bar\Psi (\x,t)\Big|_{x_3=0},
\label{mean}
\\
&&\Big(g\partial_{x_3}-\omega^2\Big)\tilde\Psi(\x,t)\Big|_{x_3=0}
\simeq g\tilde\chi\bar\Psi(\x,t)\Big|_{x_3=0}.
\label{fluct}
\eeq

From Eq. (\ref{mean}), we get the dispersion relation in the ice covered region
\beq
kg\simeq (1+\bar\chi/k)\omega^2,
\label{k_ice}
\eeq
where $1+\bar\chi/k\simeq 1+\bar\chi_0/k-\im\bar\chi_1$ plays the role of (average)
refractive index of the medium. For $\bar\chi=0$ the dispersion relation for gravity waves
in ice-free water, $kg=\omega^2$, is recovered.

\subsection{Fluctuating component}
The fluctuating component of the far-field $\tilde\Psi$ contains the diffused waves. We
can solve Eq. (\ref{fluct}) by 
requiring in the near and far field, equality of the values of the energy flux which would be generated by each element $\d A$ of the ice field in isolation.

Let us indicate with $\delta_{\d A,n}$ and with $\tilde\balpha_{\d A,n}=\tilde\bchi_n\d A$ the scattering
phase and the polarizability fluctuation of the area element 
 in $\x'$. Indicate with $\hat\alpha_{\d A,n}(\x',\varphi)=
y^{-n}\tilde\alpha^{ij\ldots}_{\d A,n}(\x')y_iy_j\ldots=\hat\alpha_{\d A,n}(\x')\cos(n\varphi)$,
the components of $\tilde\balpha_{\d A,n}$ in the direction $\y=\x-\x'$ and adopt analogous
definitions for $\tilde\chi_{\d A,n}$ and $\hat\chi_{\d A,n}$.
The value of the energy flux in the near-field region is obtained by combining
Eqs. (\ref{J}), (\ref{phi_slow}) and (\ref{phi_in,f}):
\beq
I_{\phi_{\d A,n}}&=&\im(2-\delta_{n0})
x_\perp\rho\omega\int_0^{2\pi}\d\varphi\, \Big\{\phi_{\d A,reg,n}
\int_{-\infty}^0\partial_\perp\phi^*_{\d A,sing,n}\d x_3-c.c.\Big\}_{x_\perp=0}
\nonumber
\\
&\simeq&(2n+1)\im^n\rho\omega\delta_{\d A,n}\hat\alpha^*_{\d A,n}|\bar\Psi_0|^2
\Big[x_\perp^{2+n}J_n(kx_\perp)\int_{-\infty}^0
\frac{\d x_3}{(x_\perp^2+x_3^2)^{n+3/2}}\Big]_{x_\perp=0}
\nonumber
\\
&=&\frac{(2n+1)\sqrt{\pi}}{2^{n+1}\Gamma(n+3/2)}\rho\omega 
(\im k)^n\delta_{\d A,n}\hat\alpha^*_{\d A,n}|\bar\Psi_0|^2,
\label{I in}
\eeq
where the $\delta_{n0}$ in the first line of the equation is a Kronecker delta, and
use has been made of the expression for small values of the argument of the 
Bessel function $J_n(z)\simeq (z/2)^n/\Gamma(n+1)$ \cite{abramowitz}.
The same flux evaluated in the far-field region reads, from Eq. (\ref{I}),
\beq
I_{\tilde\Psi_{\d A,n}}=\frac{4\rho\omega|\bar\Psi_0|^2}{k}\delta_{\d A,n}^2.
\label{I out}
\eeq
Requiring equality of the expressions in Eqs. (\ref{I in}) and (\ref{I out}) yields then
\beq
\delta_{\d A,n}\simeq\frac{(2n+1)\sqrt{\pi}\im^nk^{n+1}\hat\chi^*_n}{2^{n+3}\Gamma(n+3/2)}\d A,
\label{delta_0}
\eeq
where from reality of $\delta_n$, $\hat\chi_{2n}$ must be real and $\hat\chi_{2n+1}$ 
purely imaginary.


For $L_a\gg\lambda$, 
the interference of waves generated in different regions $A$ vanishes after ensemble averaging; 
the total diffused energy is therefore the sum of the energy diffused by the individual regions.
At large distance from $A$
we can write from Eq. (\ref{phi_out})
\beq
&&\tilde\Psi_{A,n}(\x,t)\simeq\im^{n+1}(2-\delta_{0n})\bar\Psi_0(t)\ex^{kx_3}\cos(n\varphi)
\int_A\d^2x'_\perp\, 
\frac{\d\delta_n}{\d A}(\x'_\perp) H_n^\smalun(k|\x_\perp-\x'_\perp|)
\nonumber
\\
&&\qquad\simeq\im^{n+1}
(2-\delta_{0n})\bar\Psi_0(t)\ex^{kx_3}\cos(n\varphi) H_n^\smalun(kx_\perp)
\int_A\d^2x'_\perp\,\frac{\d\delta_n}{\d A}(\x'_\perp)\ex^{\im\hat\k\cdot\x'_\perp}.
\label{Psi_A}
\eeq
where
\beq
\hat\k=\k-\frac{k\x_\perp}{x_\perp},
\label{khat}
\eeq
and where the following asymptotic expression of the Hankel function
for large values of the argument has been used:
\beq
H^\smalun_n(y)\simeq\sqrt{\frac{2}{\pi|y|}}\exp\Big[\im\Big(y-\frac{(2n+1)\pi}{4}\Big)\Big].
\label{asymptotics}
\eeq
From  Eq. (\ref{delta_0}) we have $\d\delta_0/d A=k\hat\chi_0/4$ and 
$\d\delta_1/d A=\im k^2\hat\chi_1^*/4$, which we substitute into Eq. (\ref{Psi_A}) and then
into Eq. (\ref{J}). We thus get the expressions for the energy density flux
\beq
\x_\perp\cdot\bar\J_A&\simeq&\frac{\rho kA\omega}{8\pi}|\bar\Psi_0|^2X(\k,\phi),
\label{J_Psi}
\\
X(\k,\varphi)&=&X_0(\k,\varphi)+[1+\cos(2\varphi)]k^2X_1(\k,\varphi),
\eeq
where
\beq
X_n(\k,\varphi)&=&(-1)^n\int\d^2x_\perp\,\overline{\hat\chi_n(\x_\perp)\hat\chi_n(0)}
\ex^{\im\hat\k\cdot\x_\perp}
\label{chi_ell}
\eeq
is the spectrum of $\hat\chi_n$ evaluated at $\hat\k=\hat\k(\k,\varphi)$,
and where we have exploited the condition $L_A\gg\lambda$ to set $A\to\infty$ in Eq. (\ref{chi_ell}).

We finally substitute Eq. (\ref{J_Psi}) into Eq. (\ref{I}) and into Eq. (\ref{sigma}), and
get the scattering cross-section of the ice in $A$:
\beq
\sigma_A(\k)\simeq \frac{kA}{8\pi}\int_{-\pi}^{\pi}\d\varphi\ X(\k,\varphi).
\label{sigma_A}
\eeq

\section{Consequences on wave attenuation}
\label{Predictions on wave attenuation}
The energy transfer to the diffused waves produces spatial attenuation of the incident
wave with the rate
\beq
q(\k)=\frac{\sigma_A(\k)}{A}
\simeq
\frac{k}{8\pi}\int_{-\pi}^{\pi}\d\varphi\ X(\k,\varphi).
\label{q close packing}
\eeq
Since we are describing wave diffusion as the result of fluctuations in the refractive
index of the ice cover, an
important parameter is the correlation length $\lambda$ 
of the field $\tilde\chi$. We consider separately the two limits $k\lambda\ll 1$ and
$k\lambda\gtrsim 1$.

\subsection{Small wavenumbers}
For small $k\lambda$, we can carry out the $ k\to 0$ limit in 
Eq. (\ref{chi_ell}); Eq.  (\ref{q close packing}) reduces to
\beq
q(\k) \simeq
\frac{kX(\varphi)}{4},
\qquad
k\lambda\ll 1,
\label{q klambda<1}
\eeq
and if $X_1=0$, scattering is isotropic.

We can apply the results to a random distribution of pointlike scattering centers 
of strength $\delta_0=k\alpha_0/4$, $\delta_1=0$, and surface density $\bar n$. We have the Poisson 
statistics result
\beq
\overline{\tilde\chi(\x_\perp)\tilde\chi(0)}=\frac{16 \bar n\delta_0^2}{k^2}\delta(\x_\perp),
\eeq
which, by
substituting into Eqs. (\ref{sigma_A}) and (\ref{q close packing}), gives us
\beq
q(\k)=\frac{4\bar n\delta_0^2}{k}=\bar n\sigma(\k),
\qquad
\bar nR^2\ll 1,\quad k\lambda\ll 1,
\label{Poisson}
\eeq
where $\sigma=4\delta_0^2/k$ is the cross-section of the individual scattering center.
The result coincides with what would be obtained in the case of incoherent diffusion
by scattering objects at separation larger than a wavelength. The situation
is similar to that of the scattering of the light of the sun by air molecules, responsible
for blue sky, an effect which can identically be described as the result of Rayleigh
scattering by individual air molecules and  that of refractive index fluctuations peaked 
at molecular scales.

\subsection{Large wavenumbers and possibility of rollover effects}
For large $k\lambda$, scattering becomes progressively concentrated
along $\k$. To illustrate the situation, we consider the case of Gaussian isotropic 
fluctuations,
$\overline{\tilde\chi(\x_\perp)\tilde\chi(0)}=\overline{\tilde\chi^2}(\varphi)
\exp[-x_\perp^2/(2\lambda^2)]$, $\overline{\tilde\chi^2}(\varphi)\equiv
\overline{\hat\chi_0^2}-k^2[1+\cos(2\varphi)]\overline{\hat\chi_1^2}$,
corresponding to the fluctuation spectrum
\beq
X(\k,\varphi)=2\pi\lambda^2\overline{\tilde\chi^2}(\varphi)
\exp[-(k\lambda)^2(1-\cos\varphi)].
\label{X klambda>>1}
\eeq
We can verify by direct
substitution of Eq. (\ref{X klambda>>1}) into Eq. (\ref{q close packing}), 
that, for large $k\lambda$, diffusion takes place at angles 
\beq
|\varphi|\sim (k\lambda)^{-1}.
\eeq
In realistic situations, however, the incident waves are not monochromatic; what one has instead is
a distribution of waves peaked at $\e_1$, with
a finite opening angle $\varphi_{inc}$. The role of directional spreading of waves in sea ice was recognized recently in Montiel \textit{et al.} (2016).\cite{montiel16}  To obtain the attenuation of the wave train, 
it is then necessary
to consider the scattering of waves at angles $|\varphi|>\varphi_{inc}$, since smaller angles
would be associated with a redistribution of the wave energy among modes within the wave train.  
We thus replace the definition of attenuation in
Eq (\ref{q close packing}) with
\beq
q _{eff}(\k)\simeq
\frac{k}{8\pi}\int_{|\varphi|>\varphi_{inc}}\d\varphi\ X(\k,\varphi).
\label{qeff}
\eeq
We can substitute Eq.  (\ref{X klambda>>1}) into Eq.  (\ref{qeff}) and evaluate the resulting
integral for large $k$ by steepest descent. We obtain
\beq
q _{eff}(\k)\simeq\frac{\overline{\tilde\chi^2}(0)}{4k\varphi_{inc}}
\exp\Big(-\frac{(k\lambda\varphi_{inc})^2}{2}\Big),
\quad
(k\lambda)^{-1}\ll\varphi_{inc}\ll 1,
\label{asympt}
\eeq
which signals that a rollover effect is indeed present at wavenumbers
\beq
k_{roll}(\k)\sim (\lambda\varphi_{inc})^{-1}.
\label{kroll}
\eeq
We note that the effect depends on our definition of  $q_{eff}$ as a loss
of energy of waves at angles $|\varphi|<\varphi_{inc}$, without distinction of
scattered and incident components. Infinite resolution and the
ability to separate the incident and the diffused component in the wave
field at angles $|\varphi|<\varphi_{inc}$, would allow us to eliminate rollover.
Substitution of Eq. (\ref{X klambda>>1}) into Eq. (\ref{q close packing}) 
and evaluation of the resulting saddle point integral for $k\lambda\gg 1$ would yield in this case
\beq
q_{eff}(\k) \sim \lambda \overline{\tilde\chi^2}(0),
\qquad 
\varphi_{inc}\ll (k\lambda)^{-1}\ll 1,
\eeq
with just a slow down of attenuation increase at large $k$,
compared to Eq. (\ref{q klambda<1}).

We point out that in realistic situations the energy spectrum of the incident waves does not have a sharp cutoff in direction, and the definition of the parameter $\varphi_{inc}$ 
remains arbitrary; the only condition that must be satisfied is $\varphi_{inc}<\pi/2$.
This freedom in the definition of $\varphi_{inc}$ provides us with a possible test of the  role of scattering in wave attenuation,
since, if wave scattering is the mechanism underlying rollover, attenuation
of waves in the angular interval $|\varphi|<\varphi_{inc}$ should decrease in response to an increase of $\varphi_{inc}$. 
In the opposite limit $k\lambda\ll 1$, diffusion is isotropic and $q_{eff}\approx q$ irrespective of the choice of $\varphi_{inc}$.

\section{Evaluation of the susceptibility function}
\label{Evaluation of}
Determining the susceptibility functions $\bchi_n$ requires knowledge of the coupled wave-ice dynamics
at the scale of the individual ice bodies. A semi-quantitative description of the dynamics
can be obtained by modeling the ice layer as 
a continuum that resists compression both horizontally and vertically but has a
relatively lower resistance to bending. This is the hypothesis at the basis
of the mass-loading model \cite{peters50,keller53,wadhams1986seasonal}, which implies that points at the bottom and the top of the ice
layer move with identical vertical velocity. We set indeed the resistance to bending
equal to zero, thus disregarding any contribution to the dynamics from the bending
rigidity of the ice bodies and the viscous stresses in the layer.
Such contributions should probably be taken into account if the size of the ice
bodies were comparable with the wavelength.

In the present hypotheses, the vertical velocity at the bottom of the ice
layer differs by an amount
\beq
V_3(\x_\perp,t)=-h(\x_\perp)\partial_{x_3}U_3(\x,t)|_{x_3=0}
=-k^2h(\x_\perp)\bar\Psi(\x,t)|_{x_3=0},
\label{V_3}
\eeq
from its value $\bar U_3(\x,t)=
\bar U_3(\x,t)|_{x_3=0}+k^2\bar\Psi(\x,t)|_{x_3=0}h(x_\perp)$ in the absence of ice,
where $h(\x_\perp)$ is a vertical scale giving the local draft of the ice layer. From comparison with 
Eq. (\ref{monopole}) and (\ref{Robincoarse}) we get immediately
\beq
\chi_0(\x_\perp)=k^2h(\x_\perp),
\label{chi_0}
\eeq
which is the same susceptibility that would be produced by a distribution of non-interacting ice
bodies of horizontal size $R$, surface number density $n\sim R^{-2}$ and polarizability
$\alpha_0\sim k^2R^2h$. On the other hand, because of horizontal 
incompressibility, the horizontal velocity of points in the ice layer is zero, and therefore 
\beq
V_1=-U_1|_{x_3=0}.
\label{V_1}
\eeq
The bottom of the ice layer is an irregular surface with characteristic horizontal and vertical roughness
scales fixed  by $R$ and $h$. Comparison of Eqs. (\ref{V_1}) and (\ref{chi}) tells us then that
the susceptibility function has a
a dipole component
\beq
\bchi_1\sim\im CkhR^2n\e_1,
\qquad
C=\pi R^2n.
\label{chi_1}
\eeq
Since for close-packed ice $C\sim 1$, the
two contribution to susceptibility in Eqs. (\ref{chi_0}) and (\ref{chi_1}) enter 
Eq. (\ref{Robincoarse}) at the same order in $kh$.
We then substitute  Eqs. (\ref{chi_0}) and (\ref{chi_1}) 
into Eq. (\ref{k_ice}) and get the following renormalized version of the mass-loading model
\beq
kg=[1+(1+r_1)k\bar h]\omega^2,
\label{mass loading}
\eeq
where $r_1$ is a dimensionless constant, which contains information on the local geometric 
structure of the ice layer and gives the strength of the roughness contribution to dispersion.

\begin{figure}
\begin{center}
\includegraphics[draft=false,width=14cm]{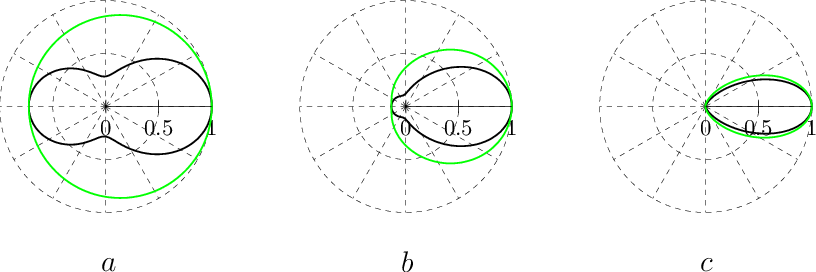}
\caption
{
Polar plot of the normalized energy current $X(\k,\varphi)/X(\k,0)$ for different values of $k$ and $r_2$: $(a)$ $k\lambda=0.4$, $(b)$ $k\lambda=1$, $(c)$ $k\lambda=2$; $r_2=1$ black line; $r_2=0$ light line (green online).
}
\label{radfig2}
\end{center}
\end{figure}

More interestingly, we can study the effect of spatial fluctuations in the layer structure on wave diffusion. 
For simplicity we continue to assume a Gaussian profile for the correlations of $\tilde\chi$.
Substituting Eqs. (\ref{chi_0}) and (\ref{chi_1}) into Eq. (\ref{X klambda>>1}) yields then
\beq
X(\k,\varphi)=2\pi k^4\lambda^2\overline{\tilde h^2}\{1+[1+\cos(2\varphi)]r_2\}
\exp[-(k\lambda)^2(1-\cos\varphi)],
\label{hat X}
\eeq
where the dimensionless constant $r_2$ plays with respect to fluctuations
a role analogous to that of $r_1$ in  Eq. (\ref{mass loading}).
The angular dependence of the function $X$ is illustrated in Fig. \ref{radfig2}. Note the increasing alignment of the energy current at $\varphi=0$ for large values of $k\lambda$.

From Eq. (\ref{asympt}), we find near the attenuation peak:
\beq
q _{eff}(\k)\simeq\frac{(1+2r_2)k^5\lambda^2\overline{\tilde h^2}}{4k\varphi_{inc}}
\exp\Big(-\frac{(k\lambda\varphi_{inc})^2}{2}\Big),
\quad
(k\lambda)^{-1}\ll\varphi_{inc}\ll 1,
\label{asympt1}
\eeq
while for small $k\lambda$ we have 
\beq
q(\k)\simeq \frac{\pi (1+r_2) k^5\lambda^2\overline{\tilde h^2}}{2},
\qquad
k\lambda\ll 1.
\label{q_h}
\eeq
We compare the result in Eq. (\ref{q_h}) with the prediction of viscous models such as the one by
Keller \cite{keller98}
\beq
q(\k) \sim \frac{k^{7/2}h\nu}{g^{1/2}},\qquad{\rm Keller},
\label{Keller}
\eeq
and 
the close-packing (CP) model \cite{desanti17}
\beq
q(\k) \sim\frac{g^{1/2}k^{5/2} h^3}{\nu},
\qquad{\rm CP},
\label{CP}
\eeq
where $\nu$ is the effective viscosity of the ice layer. A reasonable assumption in Eq. (\ref{q_h}) is that  
$\overline{\tilde h^2} \sim h^2$.
We then see that to reproduce the resulting quadratic scaling in $h$ in Eq. (\ref{q_h}), a linear
scaling in $h$ for the effective viscosity must be assumed.
The relationship is in line with both De Carolis \textit{et al.} (2021)\cite{decarolis21} and Sutherland \textit{et al.} (2019)\cite{sutherland19}, who obtained constitutive laws for $\nu=\nu(h)$ by dimensional analysis. 


We note that the wave attenuation in Eqs. (\ref{asympt1}) and (\ref{q_h}) depends on the horizontal size of the ice bodies only indirectly through the dependence of the  roughness coefficient $r_2$ on the ice concentration $C=\pi R^2n$. The absence of $R$ in the equation for $q_{eff}$ is consequence of our treating the ice layer as an almost featureless continuum.

This prompts us to look at diffusion as
the result of the interaction of the incident wave with clumps of ice of size $\lambda$. 
The correlation length $\lambda$ plays indeed a role analogous to that of the floe diameter in the
case of large floes. A simple two-dimensional wave scattering model developed in 
Wadhams (1986)\cite{wadhams1986seasonal}, in the case of floes with diameters of the order of tens of meters, proved indeed that for larger floes the rollover peak is critically dependent on the floe diameter. 

The dependence on $R$ of $q_{eff}$ resurfaces if one considers the limit of uniformly distributed ice, in which case the only surviving fluctuations are those produced by Poisson statistics at the scale of the individual ice bodies. This corresponds to making the substitution $\lambda\to R$ in Eq. (\ref{asympt1}). 
The magnitude of $q_{eff}$ is further reduced if one disregards the mutual interaction of the ice bodies, which is appropriate only for $C\ll 1$. The flow perturbation can be shown in this case to be a quadrupole field, corresponding to the polarizability of an individual body $\alpha_2\sim k^2R^4h$. From Eqs. (\ref{sigma}), (\ref{delta_0}) and (\ref{Poisson}) the resulting wave would be 
\beq
q(\k)\sim \frac{\bar n}{k}(k^3\alpha_2)^2\sim Ck^9R^6h^2,\qquad
C\ll 1,
\label{Poisson1}
\eeq
which is going to be negligible in situations of oceanographic interest.
\begin{figure}[ht]
    \centering
\includegraphics[width=1\textwidth]{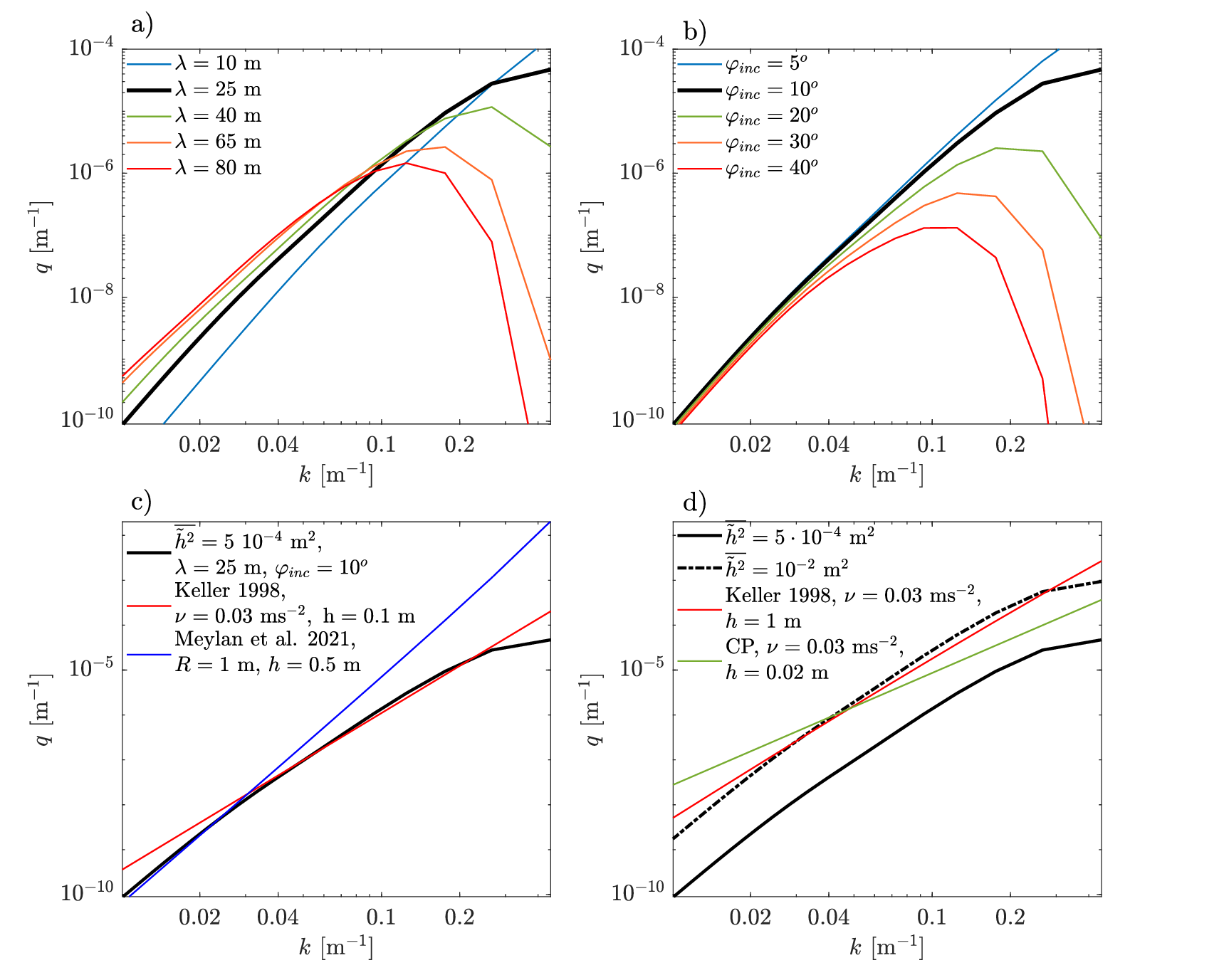}
    \caption{Trends of variation for the wave attenuation $q_{eff}$ as a function of $\lambda$ and $\varphi_{inc}$ (panels $a$ and $b$), and comparison with other viscous and scattering-based models (panels $c$ and $d$). Black solid lines corresponds to $\overline{\tilde h^2}=5\cdot 10^{-4} \rm{m}^2$, $\lambda=25$ m, $\varphi_{inc}=10^o$ and $r_2=1$ in all panels.}
    \label{models}
\end{figure}
\section{Comparison with other models}
\label{model_comparison}
We can try to put some numbers in Eqs. (\ref{hat X}) and  Eq. (\ref{qeff}), and compare the
resulting predictions on $q_{eff}$ with those by other wave attenuation models. We study the Keller's \cite{keller98} and the CP model \cite{desanti17}, and the recent
scattering-based model described in Meylan \textit{et al.} (2021)\cite{meylan2021floe}. In the last case, a computer code is provided in the reference, which we have directly utilized in our analysis.

The results are illustrated in Fig. \ref{models}, considering ocean waves with wavelengths 
ranging from $15\,{\rm m}$ to above $600\,{\rm m}$, and several values of the parameters
describing the ice layer.
In this respect, we note the different meaning taken by the ice
thickness $h$ in the different approaches: for the model in \citet{meylan2021floe}, $h$ is
the actual thickness of the floes; in \citet{keller98} and in \citet{desanti17}, it is the effective thickness
of the ice layer, while the parameter $\overline{\tilde h^2}$ 
fixes at most a fluctuation scale for $h$ in Eqs. (\ref{hat X}).

The general trends of variation for the parameters $\lambda$ and $\varphi_{inc}$ 
are illustrated in Fig. \ref{models}, panels $a$ and $b$. We note the clear rollover peaks
in attenuation that are shifted to lower wavenumbers as the two parameters $\lambda$ and
$\varphi_{inc}$ are increased.

In the scattering-based approach of Meylan \textit{et al.} (2021)\cite{meylan2021floe}, we have considered floes of radius 
$R=1\,{\rm m}$ and thickness $h=0.5\,{\rm m}$ uniformly distributed 
at the water surface (hence
$\lambda=0$, with only discrete Poisson fluctuations present).
In the case of Eq. (\ref{qeff}), we have taken values 
$\lambda=25\,{\rm m}$,  $\varphi_{inc}=10^o$, $\overline{\tilde h^2}=5\cdot 10^{-4}\,{\rm m}^2$
for the correlation length of the fluctuation in the ice layer, the opening angle of
the incident wave and the ice thickness variance, respectively, and we have set $r_2=1$ in Eq. (\ref{hat X}).

The attenuation rates predicted by the two scattering-based approaches are well-aligned only for waves longer than 200 m,
for which the flexural rigidity of 1 m radius floes is expected to be
negligible. This behavior for very long waves is consistent with computational fluid dynamics simulations that consider the heterogeneous sea ice material composition and account for the wave-ice interaction dynamics \cite{marquart2021computational}. Numerical results therein suggest that the mechanical sea ice response becomes independent of the detailed distribution of pancakes for wavenumbers smaller than 0.016 m$^{-1}$.

For wavelengths in the range $30-200\,{\rm m}$, of interest for oceanographic applications, 
the effect of the
ice cover inhomogeneity in Eq. (\ref{qeff}) becomes significant.
While scattering from uniformly distributed pancakes, as described within the approach in 
\citet{meylan2021floe}, produces attenuations almost 
proportional to $k^5$ over the whole range of $k$, taking into account long-range fluctuations
in the ice cover, as afforded by Eq. (\ref{qeff}), reduces
the exponent in the attenuation scaling to almost $7/2$.
The scaling is comparable with the one predicted by the Keller's model \cite{keller98} (see Eq. (\ref{Keller})), in the case of an effective viscosity equal to that of grease ice,\cite{newyear1999comparison}
$\nu=0.03\ \rm{ms}^{-2}$ 
 and thickness of the ice layer $h=0.1\,{\rm m}$ (red curve in Fig. \ref{models}, panel $c$).

The result is non-trivial, suggesting that scattering by pancake-ice 
in the presence of long-range fluctuations in the ice-cover
could lead to a wave damping comparable to the one due to pure viscous effects.
We note the smallness of the thickness variance  utilized, $\overline{\tilde h^2}=5\cdot10^{-4}\ \rm{m}^2$, which could rise at most to $\overline{\tilde h^2}= 1.5\cdot10^{-3}\ \rm{m}^2$ if the effect of roughness were neglected by setting $r_2=0$ in Eq. (\ref{qeff}). The
smallness of  $\overline{\tilde h^2}$
is a further indication that the effect of fluctuations in the ice cover  \textit{a priori} should not be overlooked.
The fact that the wave damping due to a fluctuation level of few centimeters is comparable with the one predicted by Keller's model could at least partially explain the high values for the effective viscosity required by viscous models to reproduce experimental data \cite{de2018ocean,cheng2017calibrating}.

The effect on $q_{eff}$ of varying $\overline{\tilde h^2}$ 
is shown in Fig. \ref{models}, panel $d$. A larger thickness 
variance in Eq. (\ref{qeff}) has the same effect on attenuation as a larger effective thickness 
of the ice layer in Keller's model. It is indeed possible to make the curves for the two 
models almost overlap
(black dash-dotted line and red continuous line in Fig. \ref{models}, panel $b$). 
The same operation does not seem to be possible with the CP model.

\section{Comparison with experiments: Arctic Sea State program data}
We compare the predictions on wave attenuation derived in the previous sections, with experimental data from wave-buoy deployed in ice fields in the Arctic.

\begin{figure}[ht]
    \centering
    \begin{minipage}[]{0.8\columnwidth}
\includegraphics[width=\columnwidth]{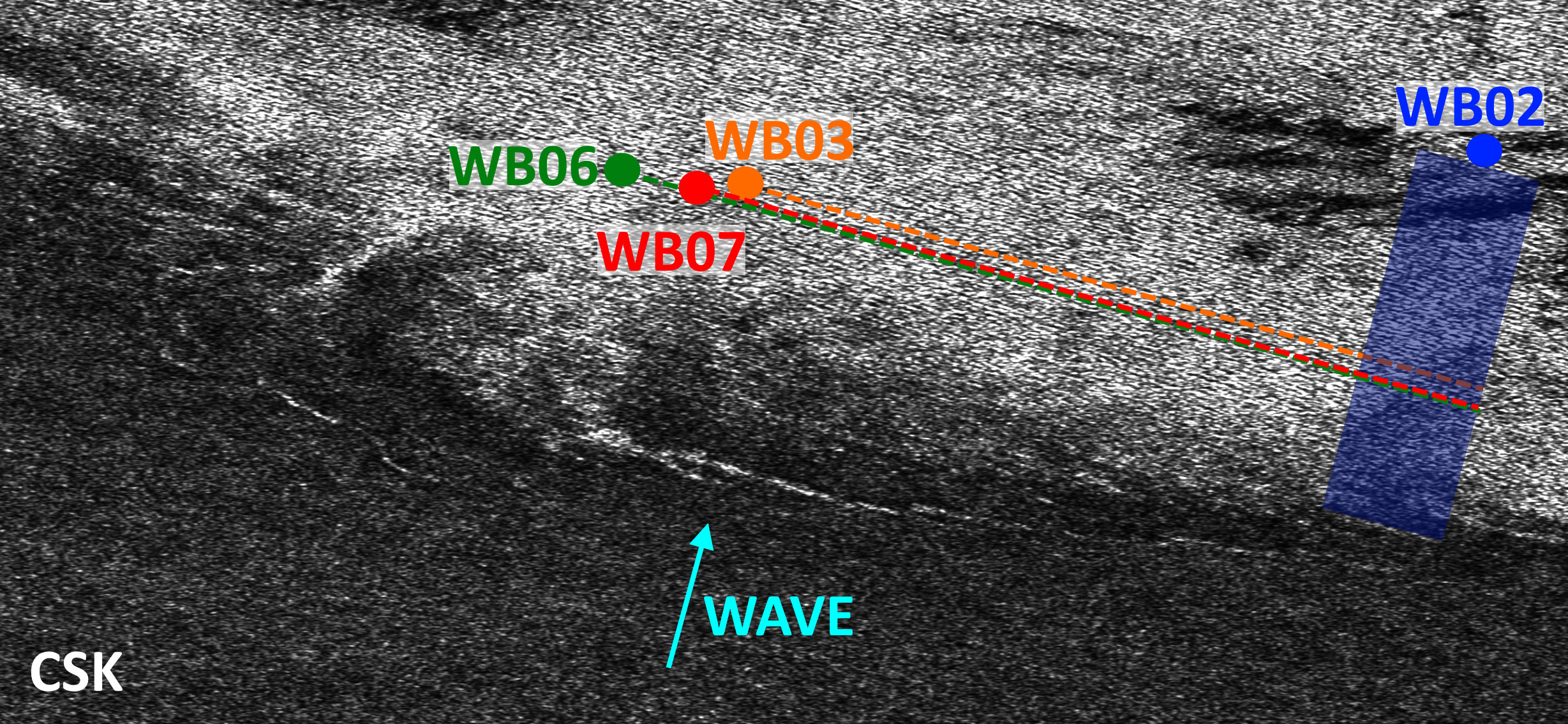}
\end{minipage}
    \begin{minipage}[]{0.8\columnwidth}
\includegraphics[width=\columnwidth]{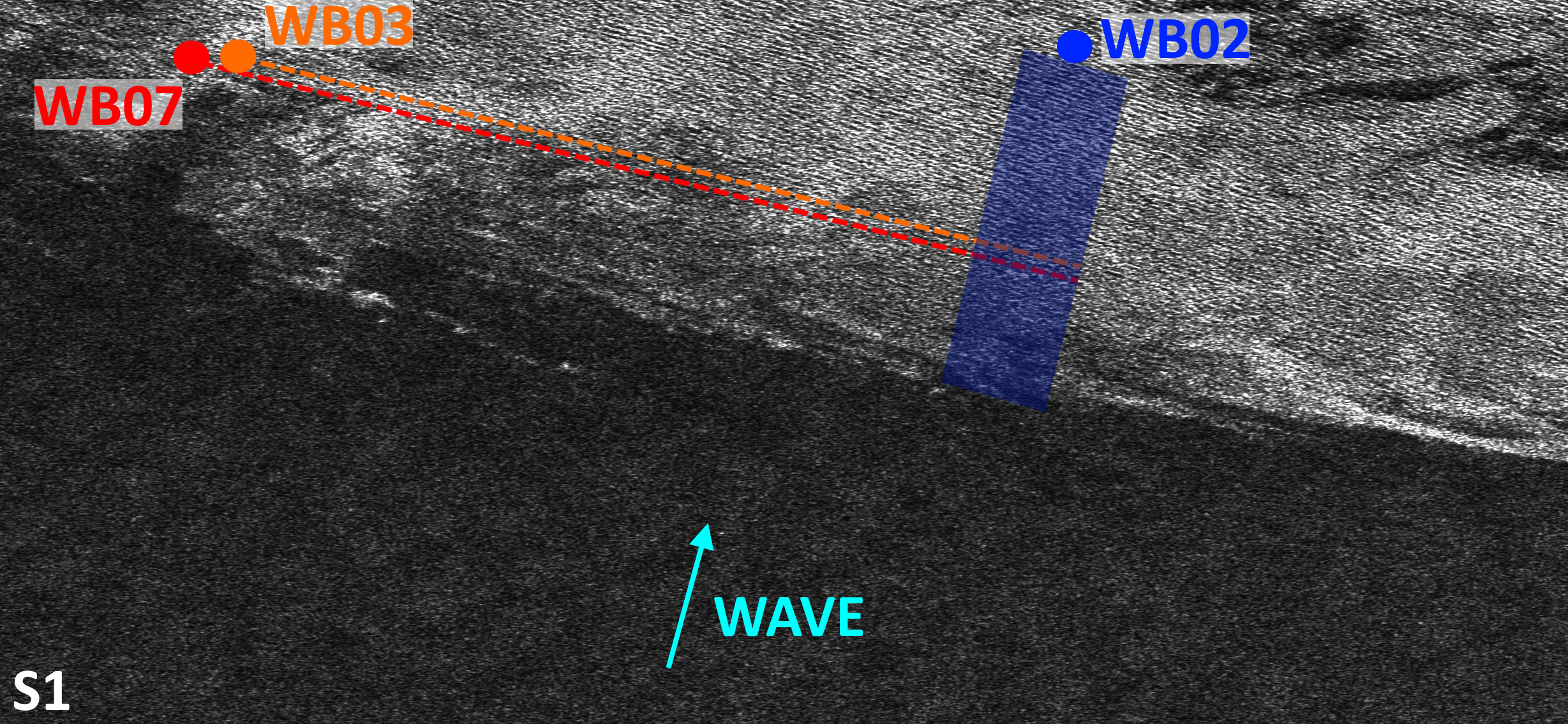}
\end{minipage}
    \caption{Top panel: Cosmo-SkyMed image acquired at 15:46 UTC.  Bottom panel: Sentinel-1A SAR image acquired at 17:20 UTC. Superimposed are the wave buoys locations at 15:30 UTC and 17:30 UTC respectivly. Cyano array indicates incoming waves peak direction. Blue rectangles highlight the area in which sea ice properties are inferred. For both images, the spatial subset [-159.6$^o$ E, -159.25$^o$ E] $\times$ [72.60$^o$ N, 72.81$^o$ N] is shown in figure.}
    \label{sarimages}
\end{figure}
The data considered in this section were collected in Autumn 2015 in the Beaufort Sea,
as part of the Arctic Sea State program \cite{thomson18}.
Wave data were supplemented by information on the properties of the ice cover (size, thickness,
concentration and composition of the ice bodies), which were supplied by the Arctic Shipborne Sea Ice Standardization Tool (ASSIST) \cite{assist} and sampling of the ice surface.
Concurrently, the sea ice extension and its temporal evolution were
monitored using satellite images acquired by Synthetic Aperture Radar (SAR) systems.

SAR observations play a crucial role in the present analysis as they provide estimates of the correlation length of the ice field.
Indeed, SAR images are generally two-dimensional maps of the Earth surface, which supply a description of the different targets in the scene with details imposed by the geometric resolution of the imaging sensor. 
The signal backscattered by the surface carries primary information on both the roughness of the surface at the scale of the impinging electromagnetic radiation and the dielectric properties of the imaged target as well \cite{franceschetti2018synthetic}. As sea ice and water have well distinct roughness and dielectric properties \cite{fu1982seasat}, SAR sensors can provide a synoptic view of the field composition \cite{toyota11}. 

Although for the MIZ a direct relation between the SAR signal
and ice properties has not been established yet, there is evidence of a correlation between ice thickness, surface roughness and the radar cross-section (i.e the SAR signal) \cite{toyota2011retrieval}. 
Information on the correlation structure of the fluctuations in the ice cover can therefore be obtained through the analysis of the correlation function of the radar cross-section.
A robust technique to estimate the correlation function from SAR is based on evaluating the inverse fast Fourier
transform (IFFT) of the power spectral density (PSD) computed from the two-dimensional radar cross section \cite{bendat2011random}. In order to remove the contribution of speckle noise at lag 0, a Butterworth filter is applied to the PSD before performing the IFFT \cite{goldfinger1982estimation, decarolis21}.

We focus on observations on the 1st of November because of the availability of two SAR images
acquired two hours apart over the area where the array of wave buoys was in operation.
The first acquisition is performed by the X-band (9.6 GHz) SAR system on-board satellite
number 3 of the Cosmo-SkyMed (CSK) constellation at 15:46 UTC; the second one by the C-band
(9.6 GHz) SAR operated by the Sentinel-1A (S1) belonging to the Copernicus mission at 17:20 UTC.
The two SAR acquisitions are shown together with the location of the wave buoys in Fig. \ref{sarimages}  and reveal a MIZ composed of grease, pancake, and small ice floes.

\begin{figure}[ht]
    \centering
    \includegraphics[width=13cm]{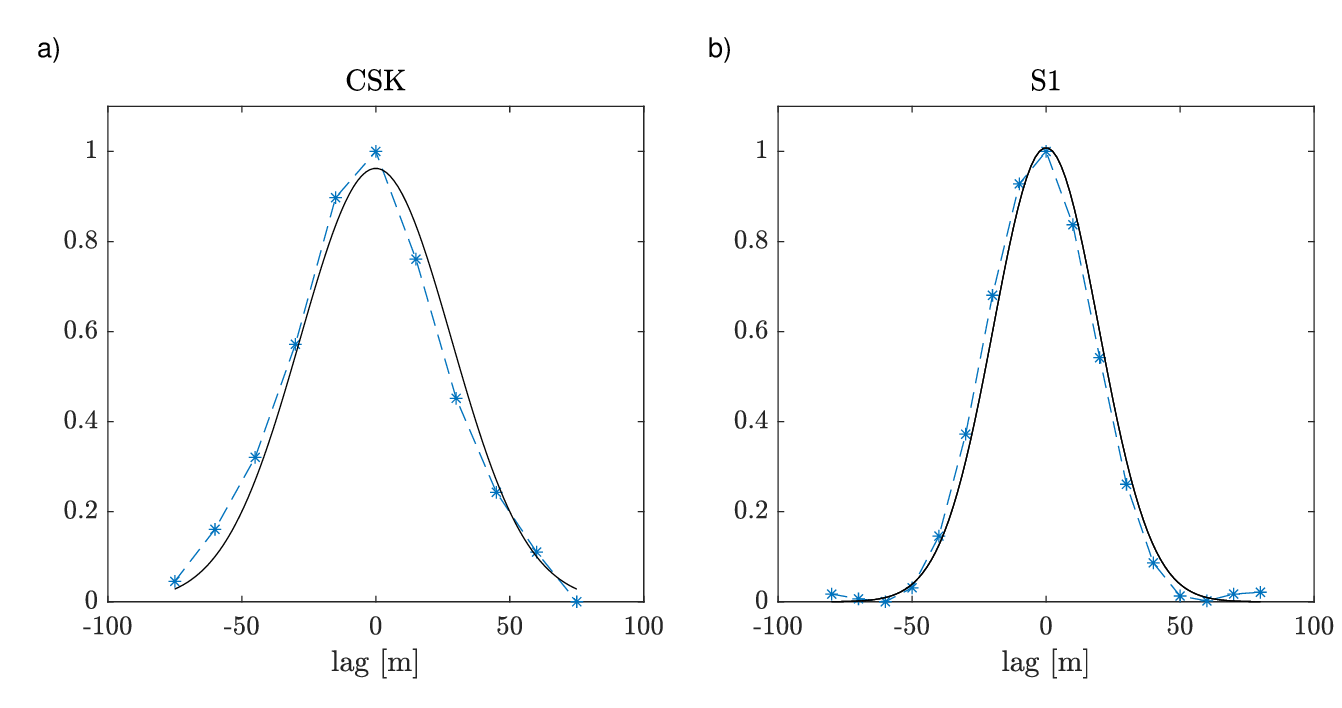}
    \caption{Normalized correlation functions 
detected on (a) CSK and (b) S1 SAR images, respectively; superimposed are the Gaussian fits.}
    \label{lambda_phi}
\end{figure}

For the blue-highligthed regions shown in Fig. \ref{sarimages}, the SAR spectral analysis is performed to determine the correlation functions along the directions of the waves coming from the open ocean ($\approx 224^o$ N) \cite{stoica2005spectral}. As shown in Fig. \ref{lambda_phi}, they are well fitted by Gaussian distributions of width $\lambda=26.83\,{\rm m}$ and  $\lambda=19.52\,{\rm m}$ 
for the CSK and S1 acquisitions, respectively.
It is worth noting that although the incoming dominant waves induce a quasi-periodic modulation of the microwave signal about 110 m long, the inferred $\lambda$ is well separated from this wavelength.

As shown in Fig. \ref{sarimages}, buoy WB02 lies farthest inside the MIZ and is therefore taken as downstream reference buoy. However, none of the buoy pairs is aligned with the dominant 
wave direction (cyan arrow). The dotted lines identify the wavefronts at the location of the buoys;
their distance to buoy WB02 defines the optical path along which wave attenuation is going to be
evaluated. The sea ice properties in the transect between the wavefronts of the upstream
buoys and buoy WB02 (blue region in Fig. \ref{sarimages}) determines the measured wave attenuation.

Let us make the hypothesis that wave diffusion is the only attenuation mechanism for the 
incident waves. We can then use the value of $\lambda$ from analysis of the SAR images
to carry out a best-fit of the buoy data on wave attenuation with Eqs. (\ref{hat X}) and
(\ref{qeff}), to retrieve the values of the remaining parameters $\varphi_{inc}$ and
$\overline{\tilde h^2}$:
\beq
\min_{\overline{\tilde h^2},\varphi_{inc}} 
\sum_i\left[q_m(\k_i)- q_{eff}(\k_i;\overline{\tilde h^2},\varphi_{inc},\lambda)\right] ^2,
\label{cost function}
\eeq
where $q_m$ are the measured data, obtained as described in Cheng \textit{et al.} (2017)\cite{cheng2017calibrating}
, and reanalyzed to clear the data from any unwanted instrumental noise energy 
\cite{thomson2020spurious}. The procedure is detailed in the Appendix.
We continue to assume $r_2=1$ in Eq. (\ref{qeff}).
\begin{figure}[ht]
\centering
\includegraphics[width=\textwidth]{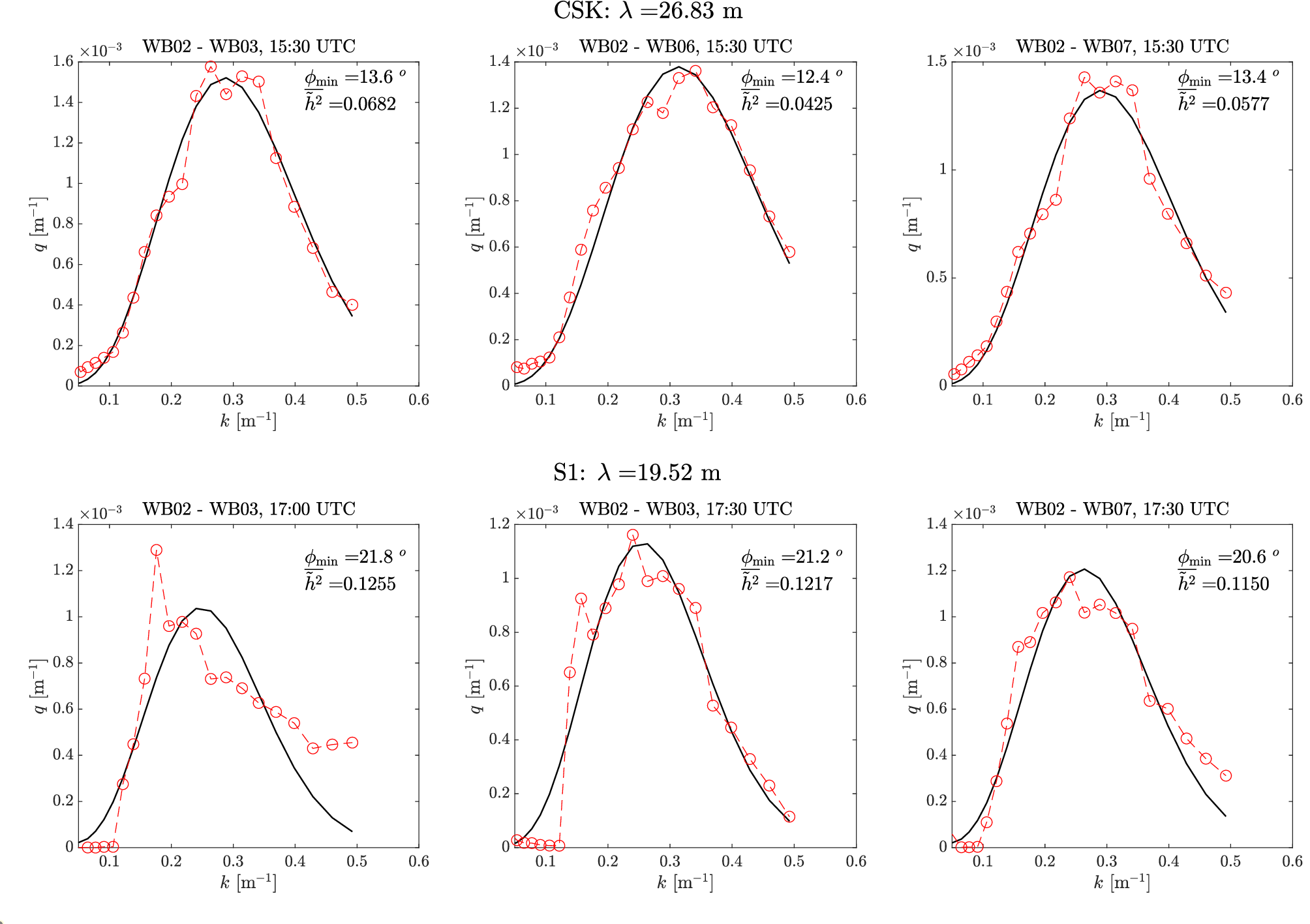}
    \caption{Examples of wave buoys' attenuation data fitted by minimizing Eq. (\ref{cost function}) for measurements concomitant with SAR acquisitions.}
    \label{qfit_1}
\end{figure}
Best-fit samples of wave attenuation for measures close to the SAR acquisitions are reported in Fig. \ref{qfit_1}. Red circles represent the wave-buoy attenuation rates $q_m$. Black lines show attenuation rates computed with Eqs. (\ref{qeff},\ref{hat X}).
The top panels show the attenuation rates measured at 15:30 UTC, and thus related to the CSK 
acquisition, while the bottom panels show the attenuation rates measured at 17:00 UTC and 17:30 UTC,
and are  related to the S1 acquisition. 

\begin{figure}[ht]
    \centering
\includegraphics[width=\textwidth]{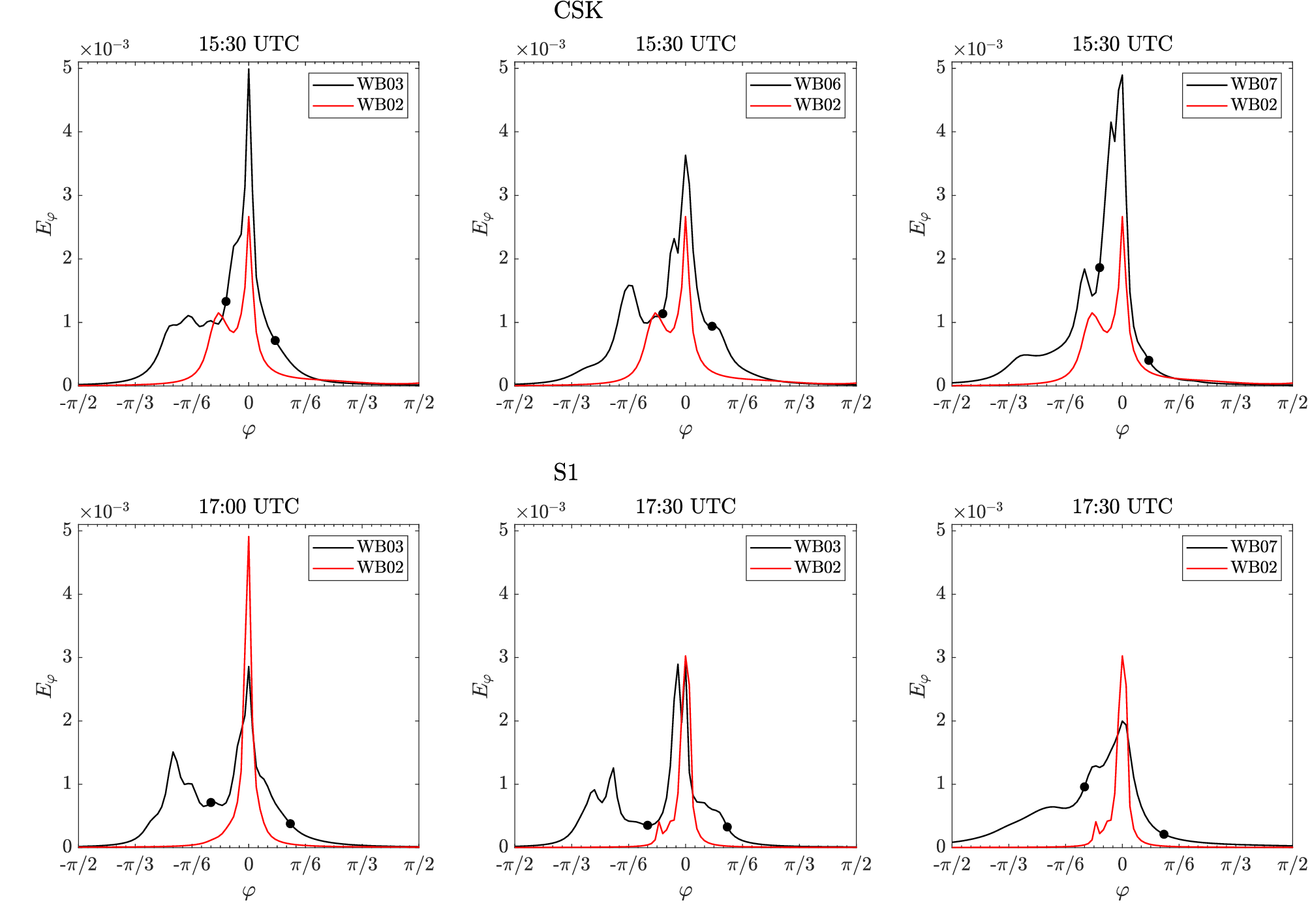}
    \caption{Wave energy spectra of the upstream buoys (black curves) and the downstream buoy (red curves) integrated over wavenumber $k$. Circles indicate the values of $\varphi_{inc}$ obtained by minimizing Eq. (\ref{cost function}). The buoy pairs reported are the same as in Figure \ref{qfit_1}.}
    \label{Ephi}
\end{figure}

In all cases, the fitting procedure can reproduce the structure of the rollover peak. 
We find in particular good agreement between the values of $\varphi_{inc}$ estimated by
the fitting procedure and the ones from analysis of the angular spectrum from 
buoy data (see Fig. \ref{Ephi}). 

The inferred values for the variance, $\overline{\tilde h^2}=0.04-0.12\,\rm{m}^2$,
instead, are large compared to the expected values of the sea-ice thickness
$h\approx 0.1\,{\rm m}$ at 16:54 UTC and $h\approx 0.15\,{\rm m}$  at 18:00 UTC,
reported in the region of interest by the Arctic Shipborne Sea Ice Standardization Tool (ASSIST) 
\cite{assist}. At the present stage, it is not clear whether the difficulty could be solved by a redefinition of $h$ and $r_2$, or is a signal of actual smallness of the diffusion contribution to wave attenuation. 

As shown in Fig. \ref{Ephi}, the inferred $\varphi_{inc}$ well separates the dominant wave from another wave packet coming further from the west. The amplitude of the second wave packet increases with time, while the energy and the opening angle of the dominant wave decrease with time. It is worth noting that the directional spectra measured at the same time by different upstream buoys show significant differences, despite the buoys locations being sufficiently tight (see Fig. \ref{sarimages}). The temporal evolution of the directional spectra of WB02 (red curves in Fig. \ref{E_Phi}), in particular, is difficult to interpret, with the peak wave energy at 17:00 UTC higher than the one of the other time stamps and the one of the upstream buoy.

\section{Conclusion}
We have studied the diffusion of gravity waves generated in a pancake-ice covered inviscid water column, with special focus on the role of random spatial variations in the thickness and the  concentration of the ice layer. 
We have modeled pancake-ice as a  horizontally inextensible, but otherwise stress-free continuum.
At the scale of the individual ice body, the dynamics is realized most simply
by assuming that the bodies are in close contact without the possibility of rafting.

Two mechanisms  in the generation of wave radiation can be identified: vertical motions of the bodies, which generate radial waves, and horizontal motions of the bodies, which generate a dipole field with lobes along the direction of propagation of the incident wave (both motions defined relative to the unperturbed wave field). The contribution from higher order multipoles appears to be negligible for bodies much smaller than a wavelength. The dipole radiation is itself negligible if
instead of a layer composed of many ice bodies, we have 
a continuous ice slab with an immersed surface of vanishing roughness.

We have evaluated the contribution to wave attenuation from diffusion in terms of the fraction of the diffused energy that is radiated at angles exceeding the opening angle $\varphi_{inc}$ of the incident wave. 

We have found that wave diffusion is stronger for waves with wavelength comparable to the correlation length
$\lambda$ of the fluctuations in the ice layer. The phenomenon is associated with a contribution to wave attenuation which is itself maximum near $\lambda$,
thus providing a mechanism for rollover similar to the one described in
Wadhams (1986)\cite{wadhams1986seasonal}, where the role of the parameter $\lambda$ was played by the radius of the individual floes. For $\lambda$  close to the observed rollover peak, it is indeed possible,
with reasonable values of the amplitude of the fluctuations in the ice thickness and concentration, to generate an attenuation of the incident waves comparable to what is observed in field experiments. One may expect that similar phenomena could play a role in wave energy harvesting 
by large assemblies of wave energy converters \cite{tokic21}. The contribution to diffusion from the discreteness of the ice bodies, instead,
is negligible for bodies much smaller than a wavelength and is further reduced if the mutual interaction of the bodies is not taken into account.

We have compared the results of the theory with buoy data from the Sikuliaq campaign in Autumn 2015 in the Arctic \cite{thomson18}, and SAR images available in that region in the same period \cite{decarolis21}. We have followed the approach in Thomson \textit{et al.} (2020)\cite{thomson2020spurious} to eliminate spurious instrumental noise contributions to the development of a rollover peak, and we have found that rollover persists in all analyzed data.  We have found that the roughness map of the SAR images are characterized by spatial fluctuations peaked in the rollover region, which  
furnishes indirect evidence to possible inhomogeneity of the ice cover at that scale. 

We have carried out a best-fit analysis of the attenuation data with the inferred value of $\lambda$, in the assumption that wave diffusion is the only mechanism at play. The resulting fluctuation levels of $h$ are somewhat large, but not to the point of dismissing the possibility of an important role of diffusion in the attenuation process. A more definite statement would require going beyond the qualitative description of the layer structure afforded in Sec.
\ref{Evaluation of}.

Attenuation by diffusion could also explain the puzzling observation that fitting experimental data by viscous models requires an effective viscosity dependent on the ice thickness. As illustrated in Sec. \ref{model_comparison}, we see in fact that the small wavenumber scaling of the attenuation obtained by a diffusion-based model can be reproduced by viscous models such as the one by Keller and the CP model only by assuming an effective viscosity $\nu$ linearly proportional to the ice thickness \cite{decarolis21}.

The analysis in this paper is based on rough modeling assumptions on the dynamics of the ice layer, which are not too different from those at the basis of the mass-loading model \cite{peters50,keller53,wadhams1986seasonal}. The average wave dynamics obtained in the present theory coincides in fact with what would be obtained using a mass-loading model. An obvious question is whether the  inextensibility hypothesis adopted in the present paper would still hold if the ice bodies in the layer interacted hydrodynamically rather than by contact forces. A numerical study of the matter is underway, following the approach detailed in \citet{kagemoto86} and followed more recently in \citet{meylan2021floe}.

\begin{acknowledgments}
This work has been supported by the Italian Piano Nazionale Ricerche in Antartide, 
(project WAMIZ, grant PNRA18/\_00109) and by the EU FP7 project ICE-ARC (grant agreement 603887).
The present work is a contribution to the Year of Polar Prediction (YOPP), a flagship activity of the Polar Prediction Project (PPP), initiated by the World Weather Research Programme (WWRP) of the World Meteorological Organisation (WMO). 
We acknowledge the WMO WWRP for its role in coordinating this international research activity. 
The CSK product used in this work was delivered by ASI (Italian Space Agency) under the ASI licence to use ID 714 in the framework of COSMO--SkyMed Open Call for Science. The Sentinel--1A used in this work was freely delivered by ESA through the Copernicus Open Access Hub (\url{https://scihub.copernicus.eu}).

\end{acknowledgments}

\appendix

\section{Data analysis procedure}
\label{appA}
The spectral noise energy $E_n(k)$ is assumed to follow the power-law $k^{-2}$ in the spectral
tail above the peak with equivalent noise height $H_n=0.1$ m. 
The attenuation rates, $q_m$, can be estimated from the measured attenuation, $q_b$, as follows:
\begin{equation}
    q_m(k)=q_b(k)+\cfrac{1}{D}\left[\cfrac{1+E_n(k)/E_D(k)}{1+E_n(k)/E_U(k)}\right]
    \label{unbiased},
\end{equation}
where $E_U$ and $E_D$ are the noise-free, omni-directional wave energy spectra of the upstream and 
the downstream buoy, respectively, and $D$ is the distance travelled by the waves between the couple of buoys. 
Note that the energy bias negligibly affected the attenuation rates $q_m$, meaning that data processing performed by \cite{cheng2017calibrating} already properly accounted for noise contamination.

\bibliography{aipsamp}

\end{document}